\documentclass[a4paper,reqno,12pt]{amsart}
\usepackage{amssymb,euscript,bbold}
\usepackage{eepic}
\usepackage{array}
\newcommand{\RR}{\mathbb{R}}
\newcommand{\U}{\mathrm{U}}
\newcommand{\SU}{\mathrm{SU}}
\renewcommand{\Sp}{\mathrm{Sp}}
\newcommand{\Spin}{\mathrm{Spin}}
\newcommand{\CC}{\mathbb{C}}
\newcommand{\EE}{\mathbb{E}}
\newcommand{\MM}{\mathbb{M}}
\newcommand{\HH}{\mathbb{H}}
\newcommand{\SO}{\mathrm{SO}}
\newcommand{\ZZ}{\mathbb{Z}}
\newcommand{\TT}{\mathbb{T}}
\newcommand{\Cl}{\mathrm{C}\ell}
\newcommand{\half}{\tfrac{1}{2}}
\newcommand{\fS}{\mathfrak{S}}
\newcommand{\M}{\mathsf{M}}
\newcommand{\eM}{\EuScript{M}}
\newcommand{\1}{\mathbb{1}}
\newcommand{\fW}{\mathfrak{W}}
\newcommand{\repre}[1]{\underline{\mathbf{#1}}}
\DeclareMathOperator{\slg}{SLAG}
\DeclareMathOperator{\clg}{\mathbb{C}LAG}
\newtheorem{dfn}{Definition}
\newtheorem{thm}{Theorem}
\newtheorem{quest}{Question}
\newtheorem{conj}{Conjecture}
\begin{document}

\title[planes, branes and automorphisms: static branes]{Planes, branes
and automorphisms\\I. Static branes}
\author[acharya]{BS Acharya}
\author[figueroa-o'farrill]{JM Figueroa-O'Farrill}
\author[spence]{B Spence}
\address[]{\begin{flushright}Department of Physics\\
Queen Mary and Westfield College\\
Mile End Road\\
London E1 4NS, UK\end{flushright}}
\email{r.acharya@qmw.ac.uk}
\email{j.m.figueroa@qmw.ac.uk}
\email{b.spence@qmw.ac.uk}
\date{\today}
\begin{abstract}
This is the first of a series of papers devoted to the
group-theoretical analysis of the conditions which must be satisfied
for a configuration of intersecting $\M5$-branes to be supersymmetric.
In this paper we treat the case of static branes.  We start by
associating (a maximal torus of) a different subgroup of $\Spin_{10}$
with each of the equivalence classes of supersymmetric configurations
of two $\M5$-branes at angles found by Ohta \& Townsend.  We then
consider configurations of more than two intersecting branes.  Such a
configuration will be supersymmetric if and only if the branes are
$G$-related, where $G$ is a subgroup of $\Spin_{10}$ contained in the
isotropy of a spinor.  For each such group we determine (a lower bound
for) the fraction of the supersymmetry which is preserved.  We give
examples of configurations consisting of an arbitrary number of
non-coincident intersecting fivebranes with fractions: $\frac{1}{32}$,
$\frac{1}{16}$, $\frac{3}{32}$, $\frac{1}{8}$, $\frac{5}{32}$,
$\frac{3}{16}$, and $\frac{1}{4}$, and we determine the resulting
(calibrated) geometry.
\end{abstract}
\maketitle

\tableofcontents

\section{Introduction}

The complete classification of supersymmetric solutions of
eleven-di\-men\-sion\-al supergravity would be an important step
towards understanding the true nature of $\M$-theory; but this seems
to be a very difficult problem.  A more manageable task seems to be
the classification of supersymmetric configurations which locally look
like intersecting $\M$-branes.  Earlier works on this topic
\cite{PT-intersecting,GKT-overlapping,Tseytlin-Mbranes,Groningen}
considered branes intersecting orthogonally.  These configurations
always preserve a fraction $\frac{1}{2^n}$ of the supersymmetry, and
all such fractions occur: $\frac{1}{4}$, $\frac{1}{8}$,
$\frac{1}{16}$, and $\frac{1}{32}$. The possibility of more
general---i.e., non-orthogonal---intersections was originally
considered in \cite{BerkoozDouglasLeigh} who noticed that branes
related by $\SU_n\subset\SO_{2n}$ transformations still preserve some
supersymmetry.  In \cite{GGPT} configurations in which the branes are
related by $\Sp_2\subset\SO_8$ transformations were shown to be dual
to Kaluza--Klein supergravity on eight-dimensional hyperk\"ahler
manifolds.  Also in \cite{BMM,BC,Costa,BLL} some of the configurations
of branes at angles were shown to be dual to branes which intersect
orthogonally.  Of course, this is impossible for configurations
preserving a fraction which is not a power of $2$.  Such
configurations were first discussed by Townsend
\cite{Townsend-Mbranes}, who initiated the classification of the
supersymmetric configurations of a pair of static $\M5$-branes at
angles, a classification completed with Ohta in \cite{OhtaTownsend};
although see also \cite{Jabbari} for some earlier but incomplete
results.  In a previous paper \cite{AFS-cali} we interpreted the
results of \cite{OhtaTownsend} in terms of calibrated geometry and
extended them to an arbitrary number of $\M5$-branes.  Similar results
in a somewhat different context have been obtained in
\cite{GibbonsPapadopoulos,GLW} also using techniques of calibrated
geometry.

The purpose of the present series of papers is to establish a
group-theoretical framework in which to phrase the analysis of these
conditions and in which to study the multiple intersection problem.
In this first paper we will consider the case of static branes, as in
the work of Ohta \& Townsend.  In a forthcoming paper \cite{AFSS},
hereafter referred to as Part~II, we will treat the general case.  We
will assign a different subgroup $G$ of $\Spin_{10}$ to each of the
configurations in \cite{OhtaTownsend}, in such a way that the
preserved supersymmetry corresponds to the spinors left invariant by
$G$, or in the case of two branes, by its maximal torus.  Many of the
groups which occur are possible holonomy groups of spin riemannian
manifolds and, by construction, all of them leave invariant a nonzero
number of spinors.  This lends further evidence to the comments in
\cite{OhtaTownsend} concerning the possible duality between these
configurations of intersecting branes and Kaluza--Klein reduction of
eleven-di\-men\-sion\-al supergravity on manifolds of reduced holonomy
\cite{GGPT}.  While we do not consider the exact brane solutions in
this paper nor discuss their Kaluza--Klein duals, we hope to return to
this question in a future publication.

This paper is organised as follows.  In the next section we recast the
analysis of Ohta \& Townsend in an intrinsically group-theoretical
fashion (see also \cite{AFS-cali}).  This will facilitate the ensuing
discussion.  In Section 3 we perform the detailed group-theoretical
analysis and associate a different group with each class of
configurations.  In Section 4 we study the case of supersymmetric
configurations involving more than two intersecting branes.  We show
that any such configuration consists of branes which are
``$G$-related'' (see below for a precise definition), where
$G\subset\Spin_{10}$ leaves a nonzero number of spinors invariant, and
given any such $G$ we determine (a lower bound for) the fraction of
the supersymmetry that a configuration of $G$-related branes will
preserve.  This yields examples of intersecting brane configurations
involving an arbitrary number of non-coincident branes which preserve
the following possible fractions of the supersymmetry: $\frac{1}{32}$,
$\frac{1}{16}$, $\frac{3}{32}$, $\frac{1}{8}$, $\frac{5}{32}$,
$\frac{3}{16}$, $\frac{1}{4}$ and $\frac{1}{2}$.  We also comment on
the geometry of some of these intersecting brane configurations.
Finally, section 5 summarises some of the open problems related to
this work.

\section{Supersymmetric pairs of $\M5$-branes at angles}

In this section we set up our notation and review the approach of Ohta
\& Townsend \cite{OhtaTownsend}.  A less detailed version of this
analysis has appeared in \cite{AFS-cali}.  Let us consider the
$\M5$-brane solution.  Let $(x^\mu)$ denote the
eleven-di\-men\-sion\-al coordinates, where $(x^0,x^1,\ldots,x^5)$ are
coordinates along the brane and $(x^6,\ldots,x^9,x^{\natural})$ are
coordinates transverse to the brane.  Far away from the brane, the
metric is asymptotically flat, so that the Killing spinors of the
supergravity solution have constant asymptotic values $\varepsilon$,
obeying
\begin{equation}\label{eq:halfsusy}
\Gamma_{012345}\, \varepsilon = \varepsilon~,
\end{equation}
where $\varepsilon$ is a real $32$-component spinor of $\Spin_{10,1}$.
We think of $\Spin_{10,1}$ as contained in the Clifford algebra
$\Cl_{1,10}$ generated by the $\Gamma_M$.  Provided we only deal with
one brane, it is possible to choose coordinates so that the brane is
stretched along these directions; but the moment we have to consider
two or more branes, particularly if they intersect non-orthogonally,
this notation becomes cumbersome, since not all branes can be
described so conveniently.  Moreover our aim in this paper is not to
analyse the global properties of branes, but their local properties at
the point of intersection.  In fact, we could be analysing
singularities in a single brane which is immersed (rather than
embedded) in the spacetime.  We will therefore recast the work of
\cite{OhtaTownsend} in terms of tangent planes at a point to the
branes themselves.

Let us fix a point $x$ in the spacetime $M$ and an orthonormal frame
$e_0,e_1,\ldots,e_9,e_\natural$ for the tangent space at $x$.  This
allows us to identify the tangent space $T_xM$ with
eleven-di\-men\-sion\-al Minkowski spacetime $\MM^{10,1}$.  We will
further decompose $\MM^{10,1} = \RR e_0 \oplus \EE^{10}$.  This
decomposition is preserved by an $\SO_{10}$ subgroup of $\SO_{10,1}$.
As in \cite{OhtaTownsend} we will restrict ourselves to configurations
for which the tangent plane to the worldvolume of a given $\M5$-brane
passing through $x$ is spanned by $e_0, v_1, \ldots, v_5$, where $v_i$
are orthonormal vectors in $\EE^{10}$.  In particular all these planes
share a common timelike direction, whence they are static relative to
one another. We will lift this restriction in Part~II.  Suppose
moreover that the brane is given the orientation defined by $e_0\wedge
v_1 \wedge \cdots \wedge v_5$.  We will therefore be able to associate
with each such brane at $x$ a 5-vector $\xi = v_1 \wedge
\cdots \wedge v_5$ in $\bigwedge^5\EE^{10}$.  Conversely, to any given
unit simple 5-vector $\xi = v_1 \wedge \cdots \wedge v_5$, we
associate an oriented 5-plane given by the span of the $v_i$.  The
condition for supersymmetry \eqref{eq:halfsusy} can be rewritten more
generally as
\begin{equation}\label{eq:fundsusy}
(e_0 \wedge \xi ) \cdot \varepsilon = \varepsilon~,
\end{equation}
where $\cdot$ stands for Clifford multiplication and where we have
used implicitly the isomorphism of the Clifford algebra
$\Cl_{1,10}$ with the exterior algebra $\bigwedge\MM^{10,1}$.  When
$\xi = e_1\wedge e_2 \wedge \cdots\wedge e_5$, equation
\eqref{eq:fundsusy} agrees with equation \eqref{eq:halfsusy}.

Now suppose that we are given two $\M5$-branes through $x$ with
tangent planes $\xi$ and $\eta$.  This configuration will be
supersymmetric if there exists a nonzero spinor $\varepsilon$ for
which
\begin{equation*}
(e_0 \wedge \xi ) \cdot \varepsilon = \varepsilon
\qquad\text{and}\qquad
(e_0 \wedge \eta ) \cdot \varepsilon = \varepsilon~.
\end{equation*}
Because $\SO_{10}$ acts transitively on the space of $5$-planes, there
exists a rotation $R$ in $\SO_{10}$ which transforms $\xi$ to $\eta$.
Because $R$ is conjugate to any given maximal torus of $\SO_{10}$,
there exists a choice of orthonormal frame $e_i$ for which $\xi =
e_1\wedge e_3 \wedge \cdots \wedge e_9$ and
\begin{equation*}
\eta = R(\theta)\xi = (\cos\theta_1 e_1 + \sin\theta_1 e_2) \wedge
\cdots \wedge (\cos\theta_5 e_9 + \sin\theta_5 e_\natural)~,
\end{equation*}
where $R(\theta)$ is the block-diagonal matrix
\begin{equation}\label{eq:rotation}
R(\theta) = 
\begin{pmatrix}
R_{12}(\theta_1) & & & & \\
& R_{34}(\theta_2) & & & \\
& & R_{56}(\theta_3) & & \\
& & & R_{78}(\theta_4) & \\
& & & & R_{9\natural}(\theta_5)
\end{pmatrix}~,
\end{equation}
each $R_{jk}(\vartheta)$ being the rotation by an angle $\vartheta$ in
the 2-plane spanned by $e_j$ and $e_k$.  The angles $(\theta_i)$ are
of course not unique, because having conjugated $R$ into a given
maximal torus, we can still act with Weyl transformations.

The Weyl group $\fW$ of $\SO_{10}$ is described as follows.  Consider
the group of permutations $\sigma$ of the ten-element set
$\{-5,\ldots,-1,1,\ldots,5\}$ such that $\sigma(-j) = -\sigma(j)$.
This group is isomorphic to the semidirect product $\fS_5 \ltimes
(\ZZ_2)^5$, where the symmetric group $\fS_5$ acts on $(\ZZ_2)^5$
interchanging the factors.  The Weyl group of $\SO_{10}$ is then the
subgroup of index $2$ consisting of even permutations.  It has order
$1920$.  Its action on the maximal torus is given by
$(\theta_1,\ldots,\theta_5) \mapsto
(\theta_{\sigma(1)},\ldots,\theta_{\sigma(5)})$ with the convention
that $\theta_{-j} = - \theta_j$.

Let $\eM$ denote the space of relative configurations of two 5-planes
in $\EE^{10}$; that is, $\eM \cong \TT/\fW$ is the quotient of the
maximal torus $\TT$ of $\SO_{10}$ by the action of the Weyl group.  If
$\theta \equiv (\theta_i)$ are any five angles, we will let $[\theta]
\in \eM$ denote their equivalence class under the action of the Weyl
group.  The subset $\eM_{\mathrm{susy}} \subset \eM$ consists of those
angles $[\theta]$ for which the intersecting brane configuration
defined by $\xi$ and $R(\theta)\xi$ preserves some supersymmetry.  In
other words, $\eM_{\mathrm{susy}}$ is the subset of $\eM$ for which
there is at least one nonzero spinor $\varepsilon$ which solves the
following equations:
\begin{equation}\label{eq:OT}
(e_0 \wedge \xi ) \cdot \varepsilon = \varepsilon
\qquad\text{and}\qquad
(e_0 \wedge R(\theta)\xi ) \cdot \varepsilon = \varepsilon~.
\end{equation}
For each point $[\theta]$ in $\eM$, let $32\nu([\theta])$ be equal to
the number of linearly independent solutions $\varepsilon$ to
\eqref{eq:OT}.  Therefore $\nu$ defines a (discontinuous) function on
$\eM$ which can be interpreted as the fraction of the supersymmetry
preserved by the configuration.  A priori $\nu$ can take any of the
values $0, \frac{1}{32}, \frac{1}{16}, \frac{3}{32}, \ldots, \half$,
but as we will see not all values actually occur.  In particular,
$\nu([\theta])=\half$ if and only if all the angles vanish, whereas
$\nu([\theta])\neq 0$ if and only if $[\theta]$ belongs to
$\eM_{\mathrm{susy}}$.

Let $\widehat R$ denote any one of the two possible lifts to
$\Spin_{10}$ of the $\SO_{10}$ rotation $R$.  Then the second equation
in \eqref{eq:OT} can be written as follows:
\begin{equation*}
\widehat{R}(\theta)\cdot (e_0 \wedge \xi) \cdot
\widehat{R}(\theta)^{-1} \cdot \varepsilon = \varepsilon~.
\end{equation*}
Using the fact that
\begin{equation}\label{eq:onlytwo}
(e_0 \wedge \xi) \cdot \widehat{R}(\theta)^{-1} =
\widehat{R}(\theta)\cdot (e_0 \wedge \xi)~,
\end{equation}
together with the first equation in \eqref{eq:OT}, we arrive at
\begin{equation}\label{eq:OT2}
\widehat{R}(\theta)^2 \cdot \varepsilon = \varepsilon~,
\end{equation}
with the same equation resulting for the other possible lift
$-\widehat{R}(\theta)$.

$\Spin_{10}$ has two complex half-spin representations $\Delta_\pm$,
obeying $\Delta_+^* \cong \Delta_-$.  Therefore their direct sum
$\Delta_+ \oplus \Delta_-$ has a real structure.  The underlying real
representation $\Delta$, defined by $\Delta\otimes_\RR \CC = \Delta_+
\oplus \Delta_-$, is the real spinor representation of $\Spin_{10,1}$
to which $\varepsilon$ belongs, whence we can think of $\varepsilon$
as a  conjugate pair of spinors, $\varepsilon = (\psi,\psi^*) \in
\Delta_+\oplus\Delta_-$.  In this way, equation \eqref{eq:OT2} simply
becomes the statement that $\psi \in \Delta_+$ is invariant under the
action of $\widehat{R}(\theta)^2 \in \Spin_{10}$.  As we shall see in
more detail below, the real and imaginary parts of each such $\psi$
give two real solutions of \eqref{eq:OT2}, but exactly one of each
such pair also obeys the first equation in \eqref{eq:OT}.  Therefore
the number of linearly independent solutions of \eqref{eq:OT} are in
one-to-one correspondence with the number of positive-chirality
spinors $\psi\in\Delta_+$ of $\Spin_{10}$ which are left invariant by
$\widehat{R}(\theta)^2$.

Notice that $\widehat{R}(\theta)^2$ is given explicitly by
\begin{equation*}
\widehat{R}(\theta)^2 = (\cos\theta_1 - \sin\theta_1 \Gamma_{12})\cdot
\cdots \cdot (\cos\theta_5 - \sin\theta_5 \Gamma_{9\natural})
\in\Cl_{1,10}~,
\end{equation*}
which is an element in the maximal torus of $\Spin_{10}$ corresponding
to the chosen maximal torus for $\SO_{10}$.  The maximal torus of
$\Spin_{10}$ acts diagonally on the space $\Delta_+$ of
positive-chirality spinors, with eigenvalues the exponentials of the
weights.  The highest weight vector of $\Delta_\pm$ is given by
$\half(1,1,1,1,\pm1)$.  All other weights in $\Delta_\pm$ are
Weyl-related to the highest weight: in particular they have
multiplicity one.  Notice also that $\lambda$ is a weight of
$\Delta_+$ if and only if $-\lambda$ is a weight of $\Delta_-$.  Now
let $\lambda$ be a weight of $\Delta_+$ and let $\varepsilon_\lambda
\in \Delta_+$ denote the unique (up to scale) weight vector of weight
$\lambda$.  Let $\varepsilon_{-\lambda} = \varepsilon_\lambda^*$
denote the corresponding weight vector in $\Delta_-$.  Taken together,
$\varepsilon_{\pm\lambda}$ are a complex basis for $\Delta_+ \oplus
\Delta_-$.  If $\lambda = \half (\sigma_1,\sigma_2,\ldots,\sigma_5)$,
where the $\sigma_i$ are signs such that their product is positive, 
then
\begin{equation*}
\widehat{R}(\theta)^2 \cdot \varepsilon_{\pm\lambda} = \exp\left(
\pm \sqrt{-1} \sum\nolimits_i \sigma_i \theta_i
\right)\,\varepsilon_{\pm\lambda}~.
\end{equation*}

Therefore the spinors left invariant by $\widehat{R}(\theta)^2$ are in
one-to-one correspondence with the weights $\lambda$ of $\Delta_+$ for
which
\begin{equation}\label{eq:zeroweight}
\lambda\cdot\theta \equiv \sum_i \sigma_i \theta_i = 0 \pmod{2\pi}~.
\end{equation}
Equivariance under the Weyl group guarantees that if we Weyl transform
the angles $\theta$ we simply Weyl transform the solutions $\lambda$.
In particular, the fraction $\nu([\theta])$, which is $\frac{1}{32}
\times$ the number of weights $\lambda$ obeying $\lambda \cdot \theta
= 0 \pmod{2\pi}$, is a well defined function on $\eM$.

Already we can characterise the space of supersymmetric configurations
$\eM_{\mathrm{susy}}$.  Let $\theta$ be some angles satisfying
equation \eqref{eq:zeroweight} for some weight $\lambda$ of
$\Delta_+$.  This weight is in the Weyl orbit of the highest weight
$\lambda_{\mathrm{max}} = \half (1,1,1,1,1)$, hence by Weyl
equivariance there will be some angles $\theta'$, Weyl-related to
$\theta$, for which $\lambda_{\mathrm{max}} \cdot \theta' = \sum_i
\theta'_i = 0 \mod{2\pi}$.  In other words, we arrive at the following
elegant characterisation of $\eM_{\mathrm{susy}}$ \cite{OhtaTownsend}:
\begin{equation}\label{eq:msusy}
\eM_{\mathrm{susy}} = \left\{[\theta]\in\eM \left| \sum_{i=1}^5
\theta_i = 0 \pmod{2\pi}\right.\right\}~.
\end{equation}
For generic $[\theta]\in\eM_{\mathrm{susy}}$, there will be a unique
weight $\lambda$ of $\Delta_+$ which satisfies $\lambda\cdot\theta = 0
\pmod{2\pi}$.  As we now explain this configuration preserves
$\frac{1}{32}$ of the supersymmetry.

We shall find it convenient to first examine the action of
$\widehat{R}(\theta)$ on $\Delta$.  Let $e_\lambda =
\varepsilon_\lambda + \varepsilon_{-\lambda}$ and $f_\lambda =
\sqrt{-1}\left( \varepsilon_\lambda - \varepsilon_{-\lambda}\right)$
denote the real and imaginary parts of the complex weight vector
$\varepsilon_\lambda$.  The set $\{e_\lambda,f_\lambda\}$ as $\lambda$
runs over the weights of $\Delta_+$ (or equivalently $\Delta_-$, since
$e_{-\lambda} = e_\lambda$ and $f_{-\lambda} = - f_\lambda$), is a
real basis for $\Delta$.  In this basis, $\widehat{R}(\theta)^2$ is no
longer diagonal, but block-diagonal with $2\times2$ blocks.  On the
two-dimensional subspace of $\Delta$ spanned by $e_\lambda$ and
$f_\lambda$, it acts with matrix
\begin{equation*}
\begin{pmatrix}
\phantom{-}\cos\lambda\cdot\theta & \sin\lambda\cdot\theta\\
-\sin\lambda\cdot\theta & \cos\lambda\cdot\theta
\end{pmatrix}~,
\end{equation*}
whence if $\lambda$ satisfies \eqref{eq:zeroweight}, both $e_\lambda$
and $f_\lambda$ are left invariant.  It might seem as if we had two
solutions per weight, but in fact the first equation in \eqref{eq:OT}
halves the number of solutions.  To see this, notice that for our
choice of 5-plane $\xi = e_1\wedge e_3 \wedge \cdots \wedge e_9$, this
equation becomes
\begin{equation}\label{eq:OTtoo}
\Gamma_{013579}\, \varepsilon = \varepsilon~.
\end{equation}
Because $\widetilde{\Gamma} \equiv \Gamma_{013579}$ anticommutes with
the Cartan generators $\Gamma_{12}$, $\Gamma_{34}$, $\Gamma_{56}$,
$\Gamma_{78}$, and $\Gamma_{9\natural}$, it preserves the subspace
$\Delta_0 \subset \Delta$ associated with the weights $\lambda$
obeying \eqref{eq:zeroweight}.  Because $\widetilde{\Gamma}^2 = +\1$,
it decomposes $\Delta_0$ into $\Delta_0^+ \oplus \Delta_0^-$ according
to its eigenvalue.  Equation \eqref{eq:OTtoo} says that $\varepsilon$
belongs to $\Delta_0^+$.  To show that $\Delta_0^\pm$ have the same
dimension, it suffices to show that $\Gamma_0$ relates them.  Indeed,
$\Gamma_0$ commutes with the Cartan generators, so that it preserves
$\Delta_0$, and anticommutes with $\widetilde{\Gamma}$ so that it maps
$\Delta_0^+$ to $\Delta_0^-$ isomorphically.

Therefore of each pair of solutions $e_\lambda$, $f_\lambda$ of
\eqref{eq:OT2}, exactly one linear combination survives
\eqref{eq:OTtoo}.  For a generic point
$[\theta]\in\eM_{\mathrm{susy}}$, there is exactly one weight
$\lambda$ in $\Delta_+$ which satisfies \eqref{eq:zeroweight}.
Therefore generically there are two linearly independent solutions
$e_\lambda$ and $f_\lambda$ of equation \eqref{eq:OT2}, one of which
satisfies equation \eqref{eq:OTtoo}.  In other words, the
configuration with characterising angles $[\theta]$ preserves
$\frac{1}{32}$ of the supersymmetry.

As described by Ohta \& Townsend \cite{OhtaTownsend} there are other
configurations preserving a larger fraction $\nu$ of the
supersymmetry.  In the next section we examine the group theory behind
these special configurations.  In particular, we will be able to
assign a different subgroup of $\Spin_{10}$ to each such
configuration.  This ``automorphism'' group of the brane configuration
often coincides with the holonomy group of a riemannian spin manifolds
possessing parallel spinors.

\section{Group-theoretical analysis}

In this section we will show how the different supersymmetric
configurations in \cite{OhtaTownsend} correspond to different
subgroups of $\Spin_{10}$ leaving some spinor(s) invariant; but before
getting into the group-theoretical description let us summarise the
results of \cite{OhtaTownsend}.

\subsection{The results of Ohta \& Townsend}

In solving equations \eqref{eq:OT}, or equivalently the first equation
in \eqref{eq:OT} and equation \eqref{eq:OT2}, it is convenient to
label the solutions according to two parameters: the fraction $\nu$ of
the supersymmetry that the configuration preserves, and the
codimension $d$ of the intersection of the two fivebranes relative to
any one of the fivebranes.  A configuration of two coincident branes
have a five-dimensional intersection, whence its codimension is zero.
At the other extreme, a configuration of two fivebranes which only
intersect in a point has codimension 5.  In most cases, the
codimension will agree with the number of nonzero angles in the
rotation matrix $R(\theta)$.  Discrepancy can occur only if any of the
angles are equal to $\pm\pi$, in which case the planes are
antiparallel and hence coincide up to orientation.  In terms of these
labels, the solutions found in \cite{OhtaTownsend} are summarised in
Table \ref{tab:OT}.  Missing from the table are a configuration with
$d=5$ and $\nu=\frac{5}{32}$ and one with $d=4$ and
$\nu=\frac{1}{4}$.  These configurations are associated with finite
subgroups of $\Spin_4\times\Spin_6$ and $\Spin_8$ respectively, and
hence consist of branes at fixed angles.

\begin{table}[h!]
\centering
\setlength{\extrarowheight}{5pt}
\begin{tabular}{|c|>{$}l<{$}|}
\hline
Codimension $d$& \multicolumn{1}{c|}{Fractions $\nu$}\\
\hline\hline
5 & \tfrac{1}{32} \to \tfrac{1}{16} \to \tfrac{3}{32}
\to \tfrac{1}{8}\\
4 & \tfrac{1}{16} \to \tfrac{1}{8} \to \tfrac{3}{16}\\
3 & \tfrac{1}{8}\\
2 & \tfrac{1}{4}\\
0 & \tfrac{1}{2}\\[5pt] \hline
\end{tabular}
\vspace{8pt}
\caption{Fractions of supersymmetry appearing in configurations of two
$\M5$-branes at angles, in terms of the codimension of the
intersection.  Arrows indicate progressive specialisation.
\label{tab:OT}}
\end{table}

We will see that with each such solution there is associated a
subgroup $G \subset \Spin_{10}$ preserving some spinor, whose maximal
torus $\TT(G)$ contains the transformations $\widehat{R}(\theta)^2$.
As discussed above, the fraction $\nu$ is determined from
the fact that $32 \nu$ is the number of zero weights of $G$, or
equivalently singlets of $\TT(G)$, acting on the half-spinor
representation $\Delta_+$ of $\Spin_{10}$.  A solution will have an
intersection of codimension $d$ whenever $\TT(G) \subset \Spin_{2d}
\subset \Spin_{10}$.

\subsection{Some regular subgroups of $\Spin_{10}$}

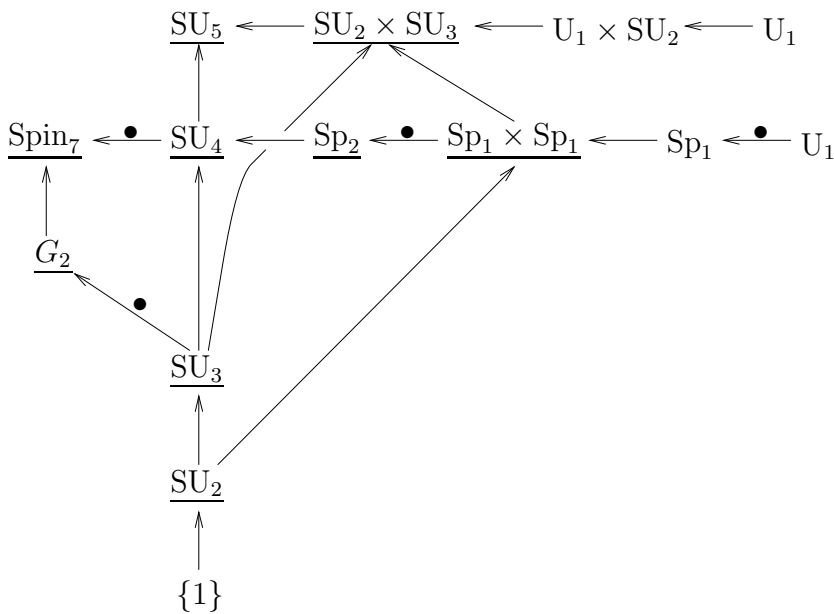
\begin{figure}[h!]
\centering
\setlength{\unitlength}{0.01in}
\begin{picture}(450,340)(-1,-10)
\put(85,300){\makebox(0,0)[lb]{$\underline{\SU_5}$}}
\path(155,310)(120,310)
\path(128,312)(120,310)(128,308)
\put(160,300){\makebox(0,0)[lb]{$\underline{\SU_2\times\SU_3}$}}
\path(280,310)(245,310)
\path(253,312)(245,310)(253,308)
\put(285,300){\makebox(0,0)[lb]{$\U_1\times\SU_2$}}
\path(355,310)(390,310)
\path(363,312)(355,310)(363,308)
\put(395,300){\makebox(0,0)[lb]{$\U_1$}}
\path(100,260)(100,300)
\path(98,292)(100,300)(102,292)
\path(265,260)(200,300)
\path(205.765, 294.104)(200,300)(207.861, 297.511)
\put(0,240){\makebox(0,0)[lb]{$\underline{\Spin_7}$}}
\path(80,250)(45,250)
\path(53,252)(45,250)(53,248)
\put(60,250){\makebox(0,0)[lb]{$\bullet$}}
\put(85,240){\makebox(0,0)[lb]{$\underline{\SU_4}$}}
\path(155,250)(120,250)
\path(128,252)(120,250)(128,248)
\put(160,240){\makebox(0,0)[lb]{$\underline{\Sp_2}$}}
\path(225,250)(190,250)
\path(198,252)(190,250)(198,248)
\put(205,250){\makebox(0,0)[lb]{$\bullet$}}
\put(230,240){\makebox(0,0)[lb]{$\underline{\Sp_1\times\Sp_1}$}}
\path(340,250)(305,250)
\path(313,252)(305,250)(313,248)
\put(345,240){\makebox(0,0)[lb]{$\Sp_1$}}
\path(375,250)(410,250)
\path(383,252)(375,250)(383,248)
\put(390,250){\makebox(0,0)[lb]{$\bullet$}}
\put(415,240){\makebox(0,0)[lb]{$\U_1$}}
\path(20,200)(20,236)
\path(22,228)(20,236)(18,228)
\path(100,140)(100,236)
\path(102,228)(100,236)(98,228)
\put(14,180){\makebox(0,0)[lb]{$\underline{G_2}$}}
\path(95,140)(35,180)
\path(42.766,177.226)(35,180)(40.547,173.898)
\put(65,160){\makebox(0,0)[lb]{$\bullet$}}
\spline(105,140)(120,230)(135,245)
\path(145,255)(190,300)
\path(182.929,295.757)(190,300)(185.757,292.929)
\put(85,120){\makebox(0,0)[lb]{$\underline{\SU_3}$}}
\path(100,80)(100,116)
\path(102,108)(100,116)(98,108)
\path(110,80)(265,236)
\path(257.943, 231.735)(265, 236)(260.78, 228.915)
\put(85,60){\makebox(0,0)[lb]{$\underline{\SU_2}$}}
\path(100,25)(100,55)
\path(102,47)(100,55)(98,47)
\put(88,3){\makebox(0,0)[lb]{$\{1\}$}}
\end{picture}
\caption{Regular subgroups of $\Spin_{10}$ associated with
intersecting brane configurations, with every arrow representing an
embedding.  Embeddings adorned with a $\bullet$ are such that the
maximal tori agree.  Underlined groups can appear as holonomy groups
of spin riemannian manifolds.}
\label{fig:groups}
\end{figure}

It is sufficient to consider only those regular subgroups
$G\subset\Spin_{10}$ which leave invariant a spinor.  A list of some
regular subgroups of $\Spin_{10}$ is given in Figure
\ref{fig:groups}.

The groups in the Figure are organised in the following fashion.  The
first row consists of subgroups of $\Spin_{10}$, the second of
subgroups of $\Spin_8$, the third of $\Spin_7$, the fourth of
$\Spin_6$, and the fifth of $\Spin_4$.  Actually $\SU_2\times\SU_3$
and its subgroups $\U_1 \times \SU_2 \supset \U_1$ are contained in
$\Spin_4 \times \Spin_6 \subset \Spin_{10}$, whereas $\Sp_1\times
\Sp_1$ and its subgroups $\Sp_1 \supset \U_1$ are contained in
$\Spin_4 \times \Spin_4 \subset \Spin_8$.  All the subgroups in the
Figure are known to preserve a spinor of $\Spin_{10}$, whence so will
their maximal tori. Indeed, as shown for example in
\cite{Bryant-spinors}, the possible isotropy groups of nonzero spinors
in $\Delta_+$ are $\SU_5$, $\Spin_7$ and their intersection $\SU_4$;
and as we can see by the embedding diagrams all the groups in the
Figure are contained in one of these. We therefore expect that the
first row should reproduce those fractions in Table \ref{tab:OT}
corresponding to configurations whose intersection has codimension
$5$, the second row should reproduce those of codimension $4$, and so
on.  The last row simply corresponds to the case of coincident branes
in which half the supersymmetry is always preserved.  Notice that
there are more groups in the Figure than fractions in Table
\ref{tab:OT}.  This discrepancy can be explained by the fact that some
of the groups in Figure \ref{fig:groups} share the same maximal torus.
Since for the case of only two branes the rotation relating them can
always be chosen to be in some maximal torus, two groups which share
the same maximal tori are indistinguishable in that their two
$\M5$-brane configurations will be identical.  This happens whenever
the embedding relates groups of equal rank.  For example, $\Spin_7$
and $\SU_4$ both have rank $3$ and have the same maximal torus in
$\Spin_8$, and so do $\Sp_2$ and $\Sp_1\times \Sp_1$ which have rank
$2$, and $\Sp_1$ and $\U_1$ which have rank $1$.  Furthermore under
the embedding $\Spin_6 \subset \Spin_7$, the maximal tori of $G_2$ and
$\SU_3$ also agree, since the maximal tori of $\Spin_7$ and of its
$\Spin_6$ subgroup are the same.  Taking these isomorphisms into
account we now see that there as many fractions in Table \ref{tab:OT}
as there are distinct maximal tori in Figure \ref{fig:groups}.  Two
features of the Figure are worth remarking.  One is the vertical
special unitary series
$\SU_5\supset\SU_4\supset\SU_3\supset\SU_2\supset\SU_1\cong\{1\}$.
The other is the horizontal spin series in the second row:
$\Spin_7\supset\Spin_6\cong\SU_4\supset\Spin_5\cong\Sp_2
\supset\Spin_4\cong\Sp_1\times\Sp_1\supset\Spin_3\cong\Sp_1
\supset\Spin_2\cong\U_1$.  In particular this series suggests that
there should be a configuration with the further subgroup
$\Spin_1\cong\ZZ_2$.  We will see how this group arises later on.

\subsection{Detailed analysis}

We now turn to the case-by-case analysis of this correspondence.  We
will simply decompose the half-spin representation $\Delta_+$ of
$\Spin_{10}$ into irreducible representations of the relevant group
$G$, and simply count the number of zero weights; that is,
$\TT(G)$-singlets.  This yields the fraction $\nu$.  The codimension
$d$ can be measured by decomposing the vector representation of
$\Spin_{10}$ and counting the number of zero weights.  If a group $G$
has $k$ zero weights in the vector representation of $\Spin_{10}$, then
the corresponding configuration will factor through $\Spin_{10-k}$,
whence it will have codimension $d = \left\lfloor \frac{10-k}{2}
\right\rfloor$.  It is actually possible to reproduce the explicit
relations on the angles which were found in \cite{OhtaTownsend} by
considering the explicit embedding of $G$ in $\Spin_{10}$ and
comparing their maximal tori.  We will however refrain from doing this
here.  The results of this section are summarised in Table
\ref{tab:summary} at the end of the section.  Much use has been made
of Slansky's Physics Report \cite{Slansky} in reaching some of the
results we are about to describe.  We will therefore follow tradition
and refer to irreducible representations by their dimensions in
agreement with \cite{Slansky}. In this notation, the half-spin
representation $\Delta_+$ of $\Spin_{10}$ is denoted $\repre{16}^*$,
whereas the vector representation is $\repre{10}$.  One small
notational disagreement worth mentioning concerns the symplectic
groups: we call $\Sp_n$ what in \cite{Slansky} would be $\Sp_{2n}$.
In our conventions, $\Sp_1 \cong \SU_2$.

\subsection{Pointlike intersections}

We now describe codimension by codimension the possible groups
responsible for the different configurations in \cite{OhtaTownsend}.
We start with codimension $d=5$ which corresponds to branes which
intersect at a point.

\subsubsection*{$\underline{\SU_5\subset\Spin_{10}}$}

The branching rules associated with this embedding are
\begin{equation*}
\repre{10} = \repre{5} \oplus \repre{5}^*\qquad\text{and}\qquad
\repre{16}^* = \repre{1} \oplus \repre{5} \oplus \repre{10}^*~.
\end{equation*}
Since we are interested in the number of singlets of the maximal torus
of $\SU_5$, we must count the number of zero weights in the weight
space decomposition of the irreducible representations appearing in
the branching rules.  This can be easily worked out for the
representations at hand.  The Dynkin labels of these $\SU_5$
representations are $\repre{5} = (1000)$, $\repre{5}^* = (0001)$ and
$\repre{10}^* = (0010)$, from which the weight decomposition can be
easily worked out.  Working in a Dynkin basis, we find:
\begin{align*}
(1000) &\to ({-}1100) \to (0{-}110) \to (00{-}11) \to (000{-}1)\\
(0001) &\to (001{-}1) \to (01{-}10) \to (1{-}100) \to ({-}1000)\\
(0010) &\to (01{-}11) \to (1{-}101) \oplus (010{-}1) \to
({-}1001) \oplus (1{-}11{-}1)\\
&\to ({-}101{-}1) \oplus (10{-}10) \to ({-}11{-}10) \to (0{-}100)~,
\end{align*}
whence we can see that the $\repre{10}$ has no zero weights, whence it
is a pointlike intersection, and that $\repre{16}^*$ has no zero
weights other than the singlet.  Thus $\nu = \tfrac{1}{32}$.

\subsubsection*{$\underline{\SU_2 \times \SU_3 \subset \Spin_4 \times
\Spin_6}$}

Let us first consider the branching rules associated with the
embedding $\Spin_4 \times \Spin_6 \subset \Spin_{10}$.
Furthermore, since $\Spin_4 \times \Spin_6$ is isomorphic to
$\SU_2 \times \SU_2 \times \SU_4$, we will work with this group
instead.  The branching rules are
\begin{equation*}
\repre{10} = (\repre{2},\repre{2},\repre{1}) \oplus
(\repre{1},\repre{1},\repre{6})\qquad\text{and}\qquad
\repre{16}^* = (\repre{2},\repre{1},\repre{4}^*) \oplus
(\repre{1},\repre{2},\repre{4})~.
\end{equation*}

Under $\SU_3 \subset \SU_4$ we have the following branching rules:
\begin{equation*}
\repre{6} = \repre{3} \oplus \repre{3}^*\qquad
\repre{4} = \repre{1} \oplus \repre{3}\qquad\text{and}\qquad
\repre{4}^* = \repre{1} \oplus \repre{3}^*~.
\end{equation*}

Of the three possible embeddings $\SU_2 \subset \SU_2 \times
\SU_2$, the diagonal embedding would have a singlet in the
$\repre{10}$ since $\repre{2}\otimes\repre{2} = \repre{3} \oplus
\repre{1}$, hence we must embed into the left factor or into the
right.  Under $\SU_2^L\times \SU_3 \subset \Spin_{10}$, we find
\begin{align*}
\repre{10} &= 2\,(\repre{2},\repre{1}) \oplus
(\repre{1},\repre{3}) \oplus
(\repre{1},\repre{3}^*)\\
\repre{16}^* &= 2\, (\repre{1},\repre{1}) \oplus
(\repre{2},\repre{1}) \oplus 2\,(\repre{1},\repre{3}) \oplus
(\repre{2},\repre{3}^*)~;
\end{align*}
whereas under $\SU_2^R \times \SU_3 \subset \Spin_{10}$, we
find
\begin{align*}
\repre{10} &= 2\,(\repre{2},\repre{1}) \oplus
(\repre{1},\repre{3}) \oplus
(\repre{1},\repre{3}^*)\\
\repre{16}^* &= 2\, (\repre{1},\repre{1}) \oplus
(\repre{2},\repre{1}) \oplus 2\,(\repre{1},\repre{3}^*) \oplus
(\repre{2},\repre{3})~.
\end{align*}
It follows that in either of the two cases, the $\repre{10}$ has no
zero weights, whereas the $\repre{16}^*$ has precisely two, coming
from the singlets.  In summary, this is a pointlike intersection with
$\nu=\frac{1}{16}$.

\subsubsection*{$\underline{\U_1\times\SU_2 \subset \Spin_4\times
\Spin_6}$}

This group $\U_1 \times \SU_2$ is actually a subgroup of
$\SU_2\times \SU_3$.  As discussed above there are two such
subgroups of $\Spin_4 \times \Spin_6$ depending on how the $\SU_2$
embeds in $\Spin_4$.  Either of the two cases yields the same
results, so we will choose to work with $\SU_2^L \times \SU_3$.
There are many conjugacy classes of $\U_1 \times \SU_2$ subgroups of
this group, but only one will give rise to a pointlike intersection
with $\nu=\frac3{32}$.  Consider the maximal subgroup $\SU_2 \times
\U_1$ of $\SU_3$.  Under $\SU_2^L\times \SU_2 \times \U_1 \subset
\SU_2^L \times \SU_3$, the $\repre{10}$ and $\repre{16}^*$ of
$\Spin_{10}$ break up as
\begin{align*}
\repre{10} &= 2\, (\repre{2},\repre{1})_0 \oplus
(\repre{1},\repre{1})_{-2} \oplus (\repre{1},\repre{2})_1
\oplus (\repre{1},\repre{1})_2 \oplus (\repre{1},\repre{2})_{-1}\\
\repre{16}^* &= 2\, (\repre{1},\repre{1})_0 \oplus
(\repre{2},\repre{1})_0 \oplus 2\,(\repre{1},\repre{1})_{-2} \oplus
2\,(\repre{1},\repre{2})_1 \oplus (\repre{2},\repre{1})_2
\oplus (\repre{2},\repre{2})_{-1}~.
\end{align*}
Because we do not desire any zero weights in the $\repre{10}$, the
extra zero weights in the $\repre{16}^*$ must come out of
representations which do not appear in the $\repre{10}$, namely
$(\repre{2},\repre{1})_2 \oplus (\repre{2},\repre{2})_{-1} \subset
\repre{16}^*$.

The $\U_1\times \SU_2$ subgroup of interest is built as follows.
The $\SU_2$ factor is the same as the $\SU_2$ factor in
$\SU_2^L \times \SU_2 \times \U_1$, whereas the $\U_1$ factor
will be embedded into $\U_1^L \times U_1$ in such a way that, if a
representation has weights $(\alpha,\beta)$ relative to $\U_1^L
\times \U_1$, it will have weight $2\alpha + \beta$ relative to the
$\U_1$ of interest.  Therefore under this $\U_1 \times \SU_2$ the
branching rules are
\begin{align*}
\repre{10} &= 3\, \repre{1}_{-2} \oplus
3\, \repre{1}_2 \oplus \repre{2}_1 \oplus \repre{2}_{-1}\\
\repre{16}^* &= 3\, \repre{1}_0 \oplus
\repre{1}_2 \oplus 3\, \repre{1}_{-2} \oplus
3\,\repre{2}_1 \oplus \repre{1}_4 \oplus \repre{2}_{-3}~.
\end{align*}
It is evident that there are no zero weights in the $\repre{10}$ but
there are three in the $\repre{16}^*$, whence, as advertised, this is
a pointlike intersection with $\nu=\frac3{32}$.  In order to
specialise this configuration further without decreasing the
codimension, it is necessary to find a subgroup of $\U_1\times\SU_2$
which has new singlets in the $\repre{16}^*$ but not in the
$\repre{10}$.  This means that this subgroup must break $\repre{1}_4
\oplus \repre{2}_{-3}$ but not any other other subrepresentations.

\subsubsection*{$\underline{\U_1 \subset  \Spin_4\times \Spin_6}$}

This $\U_1$ subgroup of the $\U_1\times\SU_2$ group discussed above
is such that the subrepresentation $\repre{2}_{-3}\subset\repre{16}^*$
has a singlet.  In other words, under this $\U_1$, a representation
with weights $(\alpha,\beta)$ under $\U_1\times \SU_2$ will have
weight $3\alpha+\beta$.  We can therefore read off the branching rules
from those above:
\begin{align*}
\repre{10} &= 4\, [-2] \oplus 4\, [2] \oplus [4] \oplus [-4]\\
\repre{16}^* &= 4\, [0] \oplus [2] \oplus 6\, [-2] \oplus
4\, [4] \oplus [-6]~.
\end{align*}
This configuration has $d=5$ and $\nu=\frac18$.  The subgroup $\ZZ_6$
defined as the kernel of the representation $\repre{1}_{-6}$ can be
shown to yield a pointlike intersection with $\nu=\frac5{32}$.

\subsection{Stringlike intersections}

The stringlike intersections are those for which $G$ is a subgroup of
$\Spin_8$: two such branes share a common one-dimensional subspace.
Under $\Spin_8 \subset \Spin_{10}$, the branching rules are
\begin{equation*}
\repre{10} = 2\, \repre{1} \oplus \repre{8}_{\mathbf{v}}
\qquad\text{and}\qquad
\repre{16}^* = \repre{8}_{\mathbf{s}} \oplus \repre{8}_{\mathbf{c}}~.
\end{equation*}

\subsubsection*{$\underline{\Spin_7 \subset \Spin_8}$}

There are three conjugacy classes of $\Spin_7$ subgroups of
$\Spin_8$, two of which leave the vector representation irreducible.
They can be distinguished by which one of the half-spin
representations they break.  In either case, the decomposition of the
$\repre{16}^*$ of $\Spin_{10}$ is the same, since both
$\repre{8}_{\mathbf{s}}$ and $\repre{8}_{\mathbf{c}}$ appear.  Indeed,
we have the following branchings:
\begin{equation*}
\repre{10} = \repre{8} \oplus 2\, \repre{1}
\qquad\text{and}\qquad
\repre{16}^* = \repre{1} \oplus \repre{7} \oplus  \repre{8}~.
\end{equation*}
In order to count singlets of the maximal torus, we perform a weight
decomposition of these irreducible representations.  Their Dynkin
labels are $\repre{7} = (100)$ and $\repre{8} = (001)$.  Therefore we
find the following weights:
\begin{align*}
(100) &\to ({-}110) \to (0{-}12) \to (000) \to (01{-2}) \to
(1{-}10) \to ({-}100)\\
(001) &\to (01{-}1) \to (1{-}11) \to ({-}101) \oplus (10{-}1) \to
({-}11{-}1)\\
&\to (0{-}11) \to (00{-}1)~.
\end{align*}
Therefore we see that there are two zero weights, coming from the two
singlets, in the decomposition of the $\repre{10}$ and also two zero
weights, one from the singlet and one from the $\repre{7}$ in the
decomposition of the $\repre{16}^*$.  In other words, it has
$d=4$ and $\nu=\frac{1}{16}$.

\subsubsection*{$\underline{\SU_4 \subset \Spin_8}$}

The same configuration is also associated with the group $\SU_4
\subset \Spin_7 \subset \Spin_8$, since as we already mentioned, the
maximal tori coincide.  The branchings in this case are
\begin{equation*}
\repre{10} = 2\, \repre{1} \oplus \repre{4} \oplus \repre{4}^*
\qquad\text{and}\qquad
\repre{16}^* = 2\,\repre{1} \oplus \repre{4} \oplus \repre{4}^*
\oplus \repre{6}~.
\end{equation*}
The Dynkin labels of these irreducible representations are $\repre{4}
= (100)$,  $\repre{4}^* = (001)$ and  $\repre{6} = (010)$; whence the
weight decompositions follow:
\begin{align*}
(100) &\to ({-}110) \to (0{-}11) \to (00{-}1)\\
(001) &\to (01{-}1) \to (1{-}10) \to ({-}100)\\
(010) &\to (1{-}11) \to ({-}101) \oplus (10{-}1)\to
({-}11{-}1)\to (0{-}10)~.
\end{align*}
Hence all zero weights come from the singlets: two in the $\repre{10}$
and two in the $\repre{16}^*$, yielding a stringlike intersection with
$\nu=\tfrac{1}{16}$.  In order to specialise further without
decreasing the codimension, we need to consider a subgroup
$G\subset\SU_4$ which has a singlet in the $\repre{6}$, but none in
the $\repre{4}$ or $\repre{4}^*$.

\subsubsection*{$\underline{\Sp_2 \subset \Spin_8}$}

The branching rules under $\Sp_2\subset\SU_4$ are given by
\begin{equation*}
\repre{4} = \repre{4}\qquad
\repre{4}^* = \repre{4}\qquad\text{and}\qquad
\repre{6} = \repre{1} \oplus \repre{5}~;
\end{equation*}
whence we can write down the decompositions of the $\repre{10}$ and the
$\repre{16}^*$ under $\Sp_2 \subset \Spin_{10}$:
\begin{equation*}
\repre{10} = 2\, \repre{1} \oplus 2\, \repre{4}
\qquad\text{and}\qquad
\repre{16}^* = 3\,\repre{1} \oplus 2\, \repre{4} \oplus \repre{5}~.
\end{equation*}
The Dynkin labels are $\repre{4} = (10)$ and $\repre{5} = (01)$, and
their weights are
\begin{align*}
(10) &\to ({-}11) \to (1{-}1) \to ({-1}0)\\
(01) &\to (2{-}1) \to (00) \to ({-}21) \to (0{-}1)~.
\end{align*}
Therefore there are exactly two zero weights in the $\repre{10}$,
coming from the singlets, and four zero weights in the $\repre{16}^*$,
three coming from the singlets and one from the $\repre{5}$.  Hence
we see that $d=4$ and $\nu=\frac{1}{8}$.  To specialise further (and
not increase $d$) we must get a singlet in the $\repre{5}$ but none in
the $\repre{4}$.

\subsubsection*{$\underline{\Sp_1\times \Sp_1 \subset \Spin_4\times
\Spin_4}$}

The same configuration can be obtained with the subgroup
$\Sp_1\times\Sp_1 \subset \Sp_2$.
In terms of $\SU_2
\times \SU_2 \times \SU_2 \times \SU_2 \cong \Spin_4
\times \Spin_4 \subset \Spin_8$, we have the following branching
rules:
\begin{align*}
\repre{8}_{\mathbf{v}} &= (\repre{2},\repre{2},\repre{1},\repre{1})
\oplus  (\repre{1},\repre{1},\repre{2},\repre{2})\\
\repre{8}_{\mathbf{s}} &= (\repre{1},\repre{2},\repre{1},\repre{2})
\oplus  (\repre{2},\repre{1},\repre{2},\repre{1})\\
\repre{8}_{\mathbf{c}} &= (\repre{1},\repre{2},\repre{2},\repre{1})
\oplus  (\repre{2},\repre{1},\repre{1},\repre{2})~.
\end{align*}
Since each $\Sp_1$ factor in $\Sp_1 \times \Sp_1$ belongs to a
different $\Spin_4$, there are four possible embeddings of
$\Sp_1\times \Sp_1$ in $\Spin_4\times\Spin_4$ which correspond to
stringlike intersections.  They all decompose the $\Spin_{10}$
representations in the same way:
\begin{align*}
\repre{10} &= 2\,(\repre{1},\repre{1}) \oplus 2\,
(\repre{1},\repre{2}) \oplus 2\, (\repre{2},\repre{1})\\
\repre{16}^* &= 4\, (\repre{1},\repre{1}) \oplus 2\,
(\repre{1},\repre{2}) \oplus 2\, (\repre{2},\repre{1}) \oplus
(\repre{2},\repre{2})~.
\end{align*}
All zero weights come from singlets, whence there are two in the
$\repre{10}$ and four in the $\repre{16}^*$.  In other words, this too
has $d=4$ and $\nu=\frac{1}{8}$.  As in the $\SU_4\subset\Spin_7$
case, one can show that this configuration is precisely the same as
from $\Sp_2$, since their maximal tori agree in $\Spin_8$.
Specialising further requires us to get a singlet in the
$(\repre{2},\repre{2})$.

\subsubsection*{$\underline{\Sp_1 \subset \Spin_4\times\Spin_4}$}

This $\Sp_1$ subgroup sits diagonally in the group $\Sp_1\times
\Sp_1$ treated above.  The branching rules can therefore be read off
immediately from the ones above:
\begin{equation*}
\repre{10} = 2\, \repre{1} \oplus 4\, \repre{2}
\qquad\text{and}\qquad
\repre{16}^* = 5\, \repre{1} \oplus 4\, \repre{2} \oplus \repre{3}~.
\end{equation*}
Thus there are two zero weights in the $\repre{10}$ coming from the
singlets and six zero weights in the $\repre{16}^*$, five coming from
the singlets and one from the $\repre{3}$. Hence this has $d=4$ and
$\nu=\frac{3}{16}$.

\subsubsection*{$\underline{\U_1 \subset  \Spin_4\times \Spin_4}$}

The same configuration can be obtained by considering the maximal
torus $\U_1$ of the above $\Sp_1$.  Under this subgroup, the
branching rules are
\begin{align*}
\repre{10} &= 2\, [0] \oplus 4\, [1] \oplus 4\, [-1]\\
\repre{16}^* &= 6\,[0] \oplus [2] \oplus 4\, [1] \oplus 4\, [-1]
\oplus [-2]~;
\end{align*}
whence we see that it has $d=4$ and $\nu=\frac3{16}$.  A further
reduction is possible to a subgroup $\ZZ_2\subset\U_1$ defined as the
kernel of the $[\pm2]$ representations.  This subgroup still has $d=4$
but now $\nu=\frac{1}{4}$.

\subsection{Two-dimensional intersections}

Here we have both subgroups of $\Spin_7$ and of $\Spin_6$.  Under
$\Spin_7 \subset \Spin_{10}$ we have
\begin{equation*}
\repre{10} = 3\,\repre{1} \oplus \repre{7}
\qquad\text{and}\qquad
\repre{16}^* = 2\,\repre{8}~,
\end{equation*}
whereas under $\Spin_6 \subset \Spin_{10}$ we have
\begin{equation*}
\repre{10} = 4\,\repre{1} \oplus \repre{6}
\qquad\text{and}\qquad
\repre{16}^* = 2\,\repre{4} \oplus 2\,\repre{4}^*~.
\end{equation*}

\subsubsection*{$\underline{G_2 \subset \Spin_7}$}

Under $G_2\subset\Spin_7$, the vector representation remains
irreducible whereas the spinor has a singlet:
\begin{equation*}
\repre{7} = \repre{7}
\qquad\text{and}\qquad
\repre{8} = \repre{1} \oplus \repre{7}~,
\end{equation*}
whence we obtain the following branchings for $G_2 \subset \Spin_{10}$:
\begin{equation*}
\repre{10} = 3\,\repre{1} \oplus \repre{7}
\qquad\text{and}\qquad
\repre{16}^* = 2\,\repre{1} \oplus 2\,\repre{7}~.
\end{equation*}
The Dynkin label of the $\repre{7}$ is $(01)$, whence its weight
decomposition is
\begin{equation*}
(01) \to (1{-}1) \to ({-}12) \to (00) \to (1{-}2) \to ({-}11) \to
(0{-}1)~.
\end{equation*}
Therefore we see that there are four zero weights in the $\repre{10}$,
three from the singlets and one from the $\repre{7}$ and also four
zero weights from the $\repre{16}^*$, two from the singlets and two
from the $\repre{7}$.  This then has $d=3$ and $\nu=\frac18$.  Notice
that this configuration admits no further specialisation with the same
codimension.  We can however obtain the same configuration with a
smaller group.

\subsubsection*{$\underline{\SU_3 \subset \Spin_6}$}

Under $\SU_3 \subset G_2$, we have that
\begin{equation*}
\repre{7} = \repre{1} \oplus \repre{3} \oplus \repre{3}^*~.
\end{equation*}
Therefore under $\SU_3 \subset \Spin_{10}$ we find
\begin{equation*}
\repre{10} = 4\,\repre{1} \oplus \repre{3} \oplus \repre{3}^*
\qquad\text{and}\qquad
\repre{16}^* = 4\,\repre{1} \oplus 2\,\repre{3} \oplus
2\,\repre{3}^*~.
\end{equation*}
All zero weights come from singlets and we have four in each
representation, hence this is also a membrane-like intersection with
$\nu=\frac18$.  In fact, one can show that this is precisely the same
configuration as the one from $G_2$, since under the embedding
$\Spin_6 \subset \Spin_7$, under which the respective maximal tori
agree, the maximal tori of $G_2$ and $\SU_3$ also agree.

\subsection{Three-dimensional intersections}

Intersections with codimension $d=2$ are such where the rotation
leaves three directions invariant, hence the rotation belongs to a
$\SO_4$ subgroup of $\SO_{10}$ one for which the vector representation
contains six singlets.  In terms of $\SU_2 \times \SU_2 \cong \Spin_4
\subset \Spin_{10}$, we find the following branching rules:
\begin{equation*}
\repre{10} = 6\, (\repre{1},\repre{1}) \oplus (\repre{2},\repre{2})
\qquad\text{and}\qquad
\repre{16}^* = 4\,(\repre{1},\repre{2}) \oplus
4\,(\repre{2},\repre{1})~.
\end{equation*}

\subsubsection*{$\underline{\SU_2 \subset \Spin_4}$}

Again there are two possible embeddings of $\SU_2$ in $\SU_2\times
\SU_2$ which do not have zero weights in the
$(\repre{2},\repre{2})$: the left and right embeddings.  Under either
one we see that
\begin{equation*}
\repre{10} = 6\, \repre{1} \oplus 2\,\repre{2}
\qquad\text{and}\qquad
\repre{16}^* =  8\,\repre{1} \oplus 4\,\repre{2}~.
\end{equation*}
All zero weights come from singlets, of which there are six in the
$\repre{10}$ and eight in the $\repre{16}^*$.  In other words this
configuration has $d=2$ and $\nu=\frac14$.  Again no further
specialisation is possible with the same codimension, since the same
irreducible representations appear in the decompositions of the 
$\repre{10}$ and $\repre{16}*$.

\subsection{Summary of results}

\begin{table}[h!]
\centering
\setlength{\extrarowheight}{5pt}
\begin{tabular}{|*{8}{>{$}c<{$}|}}
\hline
&&&
\multicolumn{2}{c|}{}&
\multicolumn{2}{c|}{Zero}&
\\
\multicolumn{1}{|c|}{Codimension}&
\multicolumn{1}{c|}{Group}&
\multicolumn{1}{c|}{Rank}&
\multicolumn{2}{c|}{Singlets}&
\multicolumn{2}{c|}{Weights}&
\multicolumn{1}{c|}{Fraction}\\
d&
G&
\ell&
\repre{10}&
\repre{16}^*&
\repre{10}&
\repre{16}^*&
\nu\\
\hline\hline
&\SU_5 & 4 & 0 & 1 & 0 & 1 & \tfrac{1}{32}\\
5&\SU_2\times\SU_3 & 3 & 0 & 2 & 0 & 2 & \tfrac{1}{16}\\
&\U_1\times\SU_2 & 2 & 0 & 3 & 0 & 3 & \tfrac{3}{32}\\
&\U_1 & 1 & 0 & 4 & 0 & 4 & \tfrac{1}{8}\\[5pt]
\hline
&\Spin_7 & 3 & 2 & 1 & 2 & 2 & \tfrac{1}{16}^\star\\
&\SU_4 & 3 & 2 & 2 & 2 & 2 & \tfrac{1}{16}\\
4&\Sp_2 & 2 & 2 & 3 & 2 & 4 & \tfrac{1}{8}^\star\\
&\Sp_1\times\Sp_1 & 2 & 2 & 4 & 2 & 4 & \tfrac{1}{8}\\
&\Sp_1 & 1 & 2 & 5 & 2 & 6 & \tfrac{3}{16}^\star\\
&\U_1 & 1 & 2 & 6 & 2 & 6 & \tfrac{3}{16}\\[5pt]
\hline
3&G_2 & 2 & 3 & 2 & 4 & 4 & \tfrac{1}{8}^\star\\
&\SU_3 & 2 & 4 & 4 & 4 & 4 & \tfrac{1}{8}\\[5pt]
\hline
2&\SU_2 & 1 & 6 & 8 & 6 & 8 & \tfrac{1}{4}\\[5pt]
\hline
0&\{1\} & 0 & 10 & 16 & 10 & 16 & \half\\[5pt]
\hline
\end{tabular}
\vspace{8pt}
\caption{Singlets and zero-weights in the decompositions of the
$\repre{10}$ and $\repre{16}^*$ of $\Spin_{10}$ under the different
groups in Figure \ref{fig:groups}, and fraction of supersymmetry in
the resulting configuration.  Fractions $\nu$ where $32\nu$ is not
equal to the number of singlets in the $\repre{16}^*$ have been
starred.}
\label{tab:summary}
\end{table}

In summary, to every supersymmetric configuration of two $\M5$-branes,
we have assigned a subgroup $G$ of $\Spin_{10}$ in such a way that the
preserved supersymmetry corresponds to the number of invariant spinors
under the action of (the maximal torus of) $G$.  These results are
summarised in Table \ref{tab:summary}, which contains the subgroups of
$\Spin_{10}$ discussed above, not including the finite subgroups.
We list the rank of the group as well as the number of singlets in the
vector and spinor representations, and the number of singlets of the
maximal torus, that is the zero weights.  The fraction of the
supersymmetry which is preserved can be read off from the number of
zero weights in the spinor representation, and the codimension can be
read off from the number of zero weights in the vector representation.
In some cases the fraction does not agree with the singlets of the
group, which means that the maximal torus leaves more spinors
invariant than the group itself.  In these cases there is a smaller
subgroup of $\Spin_{10}$ which shares the maximal torus.

These results do not just provide a group-theoretical backbone to the
results in \cite{OhtaTownsend}, but provide a basis for the extension
of these results to configurations involving more than two branes, to
which we now turn.

\section{Multiple intersections}

In this section we turn our attention to the case of multiply
intersecting branes.  The general case of more than two intersecting
branes is not immediately amenable to the kind of analysis we have
been discussing above.  The difficulty arises already for three
intersecting branes.  Suppose the three branes are parallel to start
with and rotate one of them away by a rotation $R_1$ and a second one
by a rotation $R_2$.  Unless $R_1$ and $R_2$ commute, they will not
belong to the same maximal torus, and hence it will be impossible to
choose a basis so that the matrices representing $R_1$ and $R_2$ will
have the general form \eqref{eq:rotation}.  In other words, we will
not be able to work only with maximal tori.  A different approach is
therefore needed.  In this section we will set up the problem, review
what is known and show that one can also associate an ``automorphism''
group with a given supersymmetric configuration, in such a way that
the codimension and the fraction of supersymmetry preserved can be
computed in terms of that group.  Strictly speaking we prove a theorem
which determines a lower bound for the fraction $\nu$ in terms of
group theory, and we conjecture, based on a growing body of evidence,
that the bound is actually saturated.  We will also comment on how
with every such group one can associate a geometry in the sense of
Harvey \& Lawson and we will mention some examples of such
geometries.

\subsection{Statement of the problem}

The problem is to characterise the supersymmetric configurations of
$m$ intersecting $\M5$-branes and determine the fraction $\nu$ of the
supersymmetry which is preserved.  At a mathematical level, this
problem can be formulated as follows.  Let $\Delta$ be a fixed
irreducible representation of the Clifford algebra $\Cl_{1,10}$.  It
is a real 32-dimensional representation which remains irreducible
under $\Spin_{10,1}\subset \Cl_{1,10}$.  Let $\xi$ be a $5$-plane in
$\bigwedge^5\EE^{10}$, and let $e_0\wedge \xi$ denote the tangent
plane to the worldvolume of an $\M5$-brane.  Let us define the
following subspace of $\Delta$:
\begin{equation*}
\Delta(\xi) \equiv \{\varepsilon\in\Delta | (e_0\wedge\xi)\cdot
\varepsilon = \varepsilon\}~.
\end{equation*}
$\Delta(\xi) \subset \Delta$ is a real 16-dimensional subspace.  If
$\eta$ is another $5$-plane, then define
\begin{equation*}
\Delta(\xi \cup \eta) \equiv \Delta(\xi) \cap \Delta(\eta)~,
\end{equation*}
and so on.  Let $\xi_1\equiv \xi,\xi_2,\ldots,\xi_m$ be $m$ $5$-planes.
We say that the configuration $\cup_{i=1}^m \xi_i$ is {\em
supersymmetric\/} if and only if
\begin{equation*}
\Delta(\cup_{i=1}^m \xi_i) = \bigcap_{i=1}^m \Delta(\xi_i) \neq
\{0\}~.
\end{equation*}
A supersymmetric configuration $\cup_{i=1}^m \xi_i$ is said to
preserve a fraction $\nu$ of the supersymmetry, where
\begin{equation*}
32 \nu = \dim \Delta(\cup_{i=1}^m \xi_i)~.
\end{equation*}
Clearly $\nu$ can only take the values $\tfrac{1}{32}, \tfrac{1}{16},
\tfrac{3}{32}, \ldots, \half$.  Two questions are fundamental.

\begin{quest}
How can one characterise the supersymmetric configurations
$\cup_{i=1}^m \xi_i$?
\end{quest}

\begin{quest}
What fraction $\nu$ of the supersymmetry is preserved by a given
supersymmetric configuration $\cup_{i=1}^m \xi_i$?
\end{quest}

Both questions have been answered in \cite{OhtaTownsend} for $m{=}2$.
In \cite{AFS-cali} (see also \cite{GibbonsPapadopoulos}) we answered
the first question for arbitrary $m$ using techniques of calibrated
geometry, a result we will presently recall, since it will be the
starting point of our analysis.  After doing that we will present a
partial answer to the second question for arbitrary $m$.

\subsection{Supersymmetric configurations of $G$-related planes}

How about the second question for $m>2$?  In this section we will
present a partial answer to this question and will conjecture a
complete answer based on computer experimentation; but first let us
briefly recall the result of \cite{AFS-cali} concerning the first
question for arbitrary $m$.  See also the recent work of
\cite{GibbonsPapadopoulos}.

If a configuration $\cup_{i=1}^m \xi_i$ is supersymmetric, then there
is at least one nonzero spinor $\varepsilon\in\Delta$ which belongs to
$\Delta(\xi_i)$ for all $i$.  As shown in \cite{AFS-cali}, this
implies that $\xi_i$ are calibrated by a $5$-form which can be
constructed from $\varepsilon$.  The nature of the form depends on the
isotropy subgroup of the spinor.  Nonzero spinors in eleven dimensions
have one of two possible isotropy subgroups: $\SU_5 \subset
\Spin_{10}$ and $\Spin_7 \ltimes \RR^9$ \cite{Bryant-spinors}.  In the
former case, the $5$-form is special lagrangian, and the planes
$\xi_i$ are special lagrangian planes.  Because the special lagrangian
grassmannian is isomorphic to $\SU_5/\SO_5$, we see that the planes
are all related to each other by $\SU_5$ transformations.  On the
other hand, the latter group intersects $\Spin_{10}$ in an $\Spin_7$
subgroup, and one can show that the $5$-form is now of the form $v^*
\wedge \Omega$, where $v \in \EE^{10}$ is a fixed vector, $v^*$ is the
dual $1$-form annihilating $v^\perp$ and such that $\langle
v^*,v\rangle = 1$, and $\Omega$ is a Cayley form on an $\EE^8 \subset
v^\perp$.  This means that each plane $\xi_i$ is of the form $v \wedge
\zeta_i$, where $\zeta_i$ are Cayley planes.  Because the Cayley
grassmannian is acted on transitively by $\Spin_7$, we see that the
$\zeta_i$, and hence the $\xi_i$, are related to each other by
$\Spin_7$ transformations.  These $5$-planes all intersect at least in
the subspace spanned by $v$, whence these configurations have
codimension $d\leq 4$.  The generic fraction $\nu$ is $\tfrac{1}{32}$
in either case, {\em unless\/} $m=2$ in which case, as we have seen,
planes related by $\Spin_7$ transformations are actually related by
$\SU_4$ transformations, and the fraction doubles.

Notice that $\SU_5$ and $\Spin_7$ contain all other subgroups in
Figure \ref{fig:groups}.  Therefore it is is conceivable that
demanding that the $\xi_i$ be related by transformations in some group
$G$, where $G\subset\SU_5$ or $G\subset\Spin_7$ (or both, in which
case $G\subset\SU_4$), one should obtain configurations with possibly
lower codimension and a higher fraction of supersymmetry.  The
codimension $d$ is given by $d = \left\lfloor \frac{10-k}{2}
\right\rfloor$, where $k$ is the number of $G$-singlets in the vector
representation $\repre{10}$ of $\Spin_{10}$.  One would expect that the
fraction $\nu$ of the supersymmetry would be similarly given by
$\tfrac{1}{32}\times{}$ the number of singlets in the $\repre{16}^*$.
Hence the degeneracies corresponding to the starred fractions in Table
\ref{tab:summary} would be lifted.  We have already seen that this is
true for $\Spin_7$, which now yields a fraction $\nu=\frac{1}{32}$.
Similarly we expect that $\Sp_2$ should yield a fraction
$\nu=\frac{3}{32}$, $\Sp_1\subset \Spin_8$ a fraction
$\nu=\frac{5}{32}$ and $G_2$ a fraction $\nu=\frac{1}{16}$.  We have
so far been unable to prove this, but we can prove that the fraction
is at least that.  We do this now, but first a definition.

Let $G(5,\EE^{10})$ denote the grassmannian of oriented $5$-planes in
$\EE^{10}$.  It is acted on transitively by $\SO_{10}$ with isotropy
$\SO_5\times\SO_5$.  A given subgroup $G\subset\Spin_{10}$ acts on
$G(5,\EE^{10})$ by restricting the action of $\SO_{10}$ to the
subgroup to which $G$ gets mapped under the canonical covering map
$\Spin_{10}\to\SO_{10}$.  We can therefore consider the decomposition
of the grassmannian into $G$-orbits.

\begin{dfn}
Let $\{\xi_i\}$ be $m$ oriented 5-planes in $\EE^{10}$.  We say that
they are {\em $G$-related\/}, if they all lie in the same $G$-orbit
and furthermore $G$ is the {\em smallest\/} such subgroup of
$\Spin_{10}$.
\end{dfn}

The results of \cite{AFS-cali} can be rephrased as saying that a
configuration $\cup_{i=1}^m \xi_i$ is supersymmetric if and only the
planes are $G$-related, where $G \subset \Spin_7$ or $G\subset\SU_5$
or both so that $G\subset\SU_4$.  Because both $\Spin_7$ and $\SU_5$
preserve a spinor, so will $G$. Let $\Delta^G \subset\Delta$ denote
the subspace of $G$-invariant spinors in $\Delta$.  Let the fraction
$\nu_G$ be defined by
\begin{equation}\label{eq:Gfraction}
\nu_G \equiv \tfrac{1}{64} \dim\Delta^G~.
\end{equation}
Equivalently, $32\nu_G$ is the number of linearly independent
$G$-singlets in the $\repre{16}^*$ (or the $\repre{16}$) of
$\Spin_{10}$, which can be read off from Table \ref{tab:summary} for
the subgroups discussed in Section 3.  We are now ready to prove the
following result.

\begin{thm}
Let $\cup_{i=1}^m \xi_i$ be a supersymmetric configuration of oriented
5-planes in $\EE^{10}$ which are $G$-related.  Then the fraction $\nu$
of the supersymmetry which is preserved obeys $\nu \geq \nu_G$.
\end{thm}

\begin{proof}
Let $\xi\equiv \xi_1$, say, be one of the planes, and let $\pi = e_0
\wedge \xi$.  Also let $\pi_i = e_0\wedge \xi_i$, for $i=1,\ldots,m$,
so that $\pi_1 = \pi$.  Because the $\xi_i$ are $G$-related, and
$G\subset\Spin_{10}$ acts trivially on $e_0$, so are the $\pi_i$.
This means that there are group elements $g_i\in G$, unique modulo the
isotropy of $\xi$, so that $\xi_i = g_i\,\xi$ and $\pi_i = g_i\,\pi$.
Now let $\varepsilon \in \Delta^G \cap \Delta(\xi)$; that is,
$\varepsilon$ is a $G$-invariant spinor which obeys $\pi \cdot
\varepsilon = \varepsilon$.  It is plain that $\varepsilon$ also obeys
$\pi_i \cdot \varepsilon = \varepsilon$ for all $i$.  Indeed,
\begin{align*}
\pi_i \cdot \varepsilon &= g_i\cdot \pi \cdot g_i^{-1} \cdot
\varepsilon\\
&= g_i \cdot \pi \cdot \varepsilon \tag{$\varepsilon\in\Delta^G$}\\
& = g_i \cdot \varepsilon \tag{$\varepsilon\in\Delta(\xi)$}\\
& = \varepsilon~, \tag{$\varepsilon\in\Delta^G$}
\end{align*}
whence $\varepsilon \in \Delta(\xi_i)$ for all $i$.  In other words,
we have shown that
\begin{equation*}
\Delta^G \cap \Delta(\xi) \subset \Delta(\cup_{i=1}^m \xi_i)~,
\end{equation*}
whence
\begin{equation}\label{eq:preineq}
\nu \equiv \tfrac{1}{32} \dim \Delta(\cup_{i=1}^m \xi_i) \geq
\tfrac{1}{32} \dim \left(\Delta^G \cap \Delta(\xi)\right)~.
\end{equation}
We will now show that $\Delta^G \cap \Delta(\xi)$ has half the
dimension of $\Delta^G$.

Because $\pi\cdot\pi = \1$, we have a decomposition
\begin{equation*}
\Delta = \Delta^+ \oplus \Delta^-
\end{equation*}
into eigenspaces of $\pi$.  Clearly $\Delta^+ = \Delta(\xi)$.  The
above decomposition allows us to decompose $\Delta^G$:
\begin{equation*}
\Delta^G = \Delta^G_+ \oplus \Delta^G_-~,
\end{equation*}
where $\Delta^G_\pm = \Delta^G \cap \Delta^\pm$.  Now consider the
action of $e_0$ on $\Delta$.  Because $G \subset \Spin_{10}$, $g \cdot
e_0 = e_0 \cdot g$ for all $g\in G$.  Therefore $e_0$ preserves
$\Delta^G$.  Furthermore, $e_0 \cdot \pi = - \pi \cdot e_0$, whence
$e_0$ maps $\Delta^+$ to $\Delta^-$, and also $\Delta^G_+$ to
$\Delta^G_-$.  Because $e_0\cdot e_0=-\1$, it is an isomorphism, and
$\Delta^G_\pm$ have the same dimension: one half the dimension of
$\Delta^G$.  In other words, $\Delta^G \cap \Delta(\xi)$ has half the
dimension of $\Delta^G$.  Together with \eqref{eq:preineq}, the
theorem follows.
\end{proof}

There is a large body of evidence which suggests that the inequality
in the Theorem is actually saturated.  The $m=2$ results described in
Section 3 support this, and so do the results of computer
experimentation.  We therefore feel confident in the validity of the
following conjecture.  It essentially asserts that there is no
accidental supersymmetry, beyond that which is guaranteed by the group
theory.

\begin{conj}
A supersymmetric configuration $\cup_{i=1}^m \xi_i$ of oriented
5-planes in $\EE^{10}$ which are $G$-related preserves a fraction
$\nu_G$ of the supersymmetry, where $\nu_G$ is given by
\eqref{eq:Gfraction}.
\end{conj}

\begin{figure}[h!]
\centering
\setlength{\unitlength}{0.01in}
\begin{picture}(360,460)(-10,-130)
\put(0,300){\makebox(0,0)[lb]{$\nu_G = \frac{1}{32}$}}
\put(180,300){\makebox(0,0)[cb]{$\SU_5$}}
\put(280,300){\makebox(0,0)[cb]{$\Spin_7$}}
\path(130,260)(170,295)
\path(162.662, 291.237)(170, 295)(165.296, 288.227)
\path(225,260)(185,295)
\path(189.704, 288.227)(185., 295.)(192.338, 291.237)
\path(235,260)(275,295)
\path(267.662, 291.237)(275., 295.)(270.296, 288.227)
\path(330,260)(290,295)
\path(294.704, 288.227)(290., 295.)(297.338, 291.237)
\put(0,240){\makebox(0,0)[lb]{$\nu_G = \frac{1}{16}$}}
\put(130,240){\makebox(0,0)[cb]{$\SU_2\times\SU_3$}}
\put(230,240){\makebox(0,0)[cb]{$\SU_4$}}
\put(330,240){\makebox(0,0)[cb]{$G_2$}}
\path(130,200)(130,235)
\path(128,227)(130,235)(132,227)
\path(230,200)(230,235)
\path(228,227)(230,235)(232,227)
\put(0,180){\makebox(0,0)[lb]{$\nu_G = \frac{3}{32}$}}
\put(135,180){\makebox(0,0)[cb]{$\U_1\times\SU_2$}}
\put(230,180){\makebox(0,0)[cb]{$\Sp_2$}}
\path(130,140)(130,175)
\path(128,167)(130,175)(132,167)
\path(230,140)(230,175)
\path(228,167)(230,175)(232,167)
\path(220,140)(140,235)
\path(143.623, 227.592)(140., 235.)(146.683, 230.169)
\path(330,140)(330,235)
\path(328,227)(330,235)(332,227)
\spline(315,140)(260,205)(235,212)
\path(225,215)(150,235)
\path(157.215, 231.006)(150., 235.)(158.245, 234.871)
\path(325,140)(240,235)
\path(243.844, 227.704)(240., 235.)(246.825, 230.372)
\put(0,120){\makebox(0,0)[lb]{$\nu_G = \frac{1}{8}$}}
\put(130,120){\makebox(0,0)[cb]{$\U_1$}}
\put(230,120){\makebox(0,0)[cb]{$\Sp_1\times\Sp_1$}}
\put(330,120){\makebox(0,0)[cb]{$\SU_3$}}
\path(230,80)(230,115)
\path(228,107)(230,115)(232,107)
\put(0,60){\makebox(0,0)[lb]{$\nu_G = \frac{5}{32}$}}
\put(230,60){\makebox(0,0)[cb]{$\Sp_1$}}
\path(230,20)(230,55)
\path(228,47)(230,55)(232,47)
\put(0,0){\makebox(0,0)[lb]{$\nu_G = \frac{3}{16}$}}
\put(230,0){\makebox(0,0)[cb]{$\U_1$}}
\path(330,-40)(330,115)
\path(328,107)(330,115)(332,107)
\path(320,-40)(240,115)
\path(241.892, 106.974)(240., 115.)(245.446, 108.808)
\put(0,-60){\makebox(0,0)[lb]{$\nu_G = \frac{1}{4}$}}
\put(330,-60){\makebox(0,0)[cb]{$\SU_2$}}
\path(230,-100)(230,-5)
\path(228,-13)(230,-5)(232,-13)
\path(240,-100)(320,-65)
\path(311.869, -66.3742)(320., -65.)(313.472, -70.0389)
\path(220,-100)(130,115)
\path(131.244, 106.848)(130., 115.)(134.934, 108.393)
\put(0,-120){\makebox(0,0)[lb]{$\nu_G = \frac{1}{2}$}}
\put(230,-120){\makebox(0,0)[cb]{$\{1\}$}}
\end{picture}
\caption{Fractions of supersymmetry associated to $G$-related planes
as a function of $G$.  Each arrow denotes an embedding.}
\label{fig:Gfractions}
\end{figure}
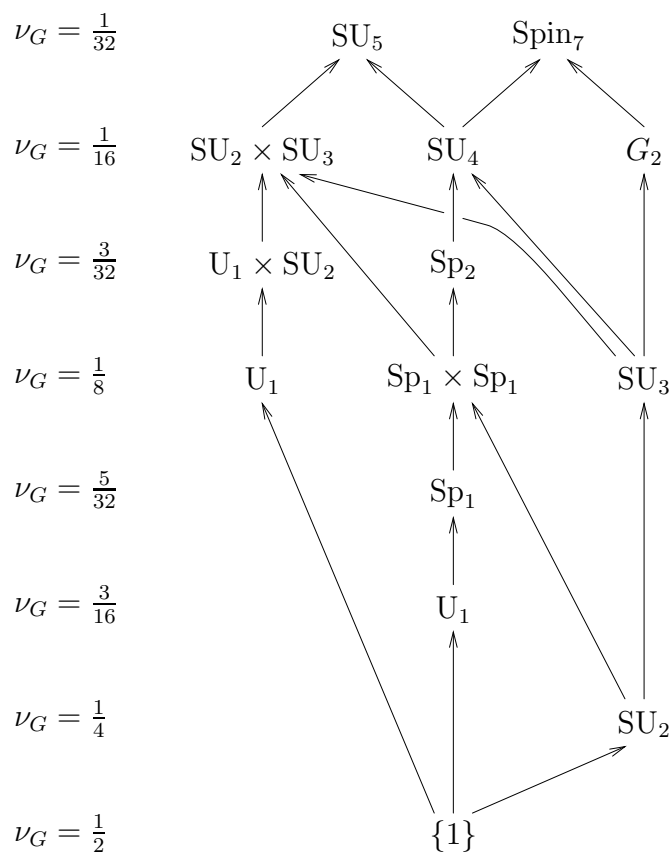

We can give many examples of configurations of an arbitrary
number of intersecting $\M5$-branes preserving a certain fraction of
the supersymmetry.  We simply choose the planes to be $G$-related
where $G \subset\Spin_{10}$ is a given subgroup of $\SU_5$ or
$\Spin_7$ (or both).  Modulo the conjecture, the fraction $\nu =
\nu_G$ can then be read off from Table \ref{tab:summary}, and in any
case the fraction will be at least $\nu_G$.  The possible fractions
are as in the $m=2$ case: $\frac{1}{32}$, $\frac{1}{16}$,
$\frac{3}{32}$, $\frac{1}{8}$, $\frac{5}{32}$, $\frac{3}{16}$,
$\frac{1}{4}$ and $\frac{1}{2}$.  These results are summarised in
Figure \ref{fig:Gfractions}.  It is worth remarking, however, that for
$m=2$, the fraction $\nu = \frac{5}{32}$ appears for branes which are
$\Gamma$-related, where $\Gamma$ is a certain $\ZZ_6$ subgroup of 
$\Spin_4\times\Spin_6\subset\Spin_{10}$.  Using this group,
configurations with $\nu=\frac{5}{32}$ would of necessity consist of a 
finite number of non-coincident branes.  In contrast, for generic $m$,
this fraction is associated to an $\Sp_1$ subgroup of $\Spin_4 \times
\Spin_4 \subset \Spin_8$, whence configurations with
$\nu=\frac{5}{32}$ consisting an arbitrary number of non-coincident
branes are possible.

\subsection{Geometry of intersecting brane configurations}

We have seen how with a given supersymmetric static configuration of
intersecting branes one can associate a Lie subgroup $G \subset
\Spin_{10}$ such that the different branes are $G$-related.  Moreover
we have conjectured a precise relation between the fraction of the
supersymmetry preserved by such a configuration and the dimension of
the space of $G$-invariant spinors.  We will now refine this
correspondence and associate with every such configuration a given
geometry: this correspondence is most clearly seen in the formalism of
calibrated geometry \cite{HarveyLawson}.  For a review of the basic
notions of calibrated geometry in the present context, see our
previous paper \cite{AFS-cali} as well as references therein.  Other
recent papers which discuss calibrated geometry in the context of
intersecting branes are \cite{GibbonsPapadopoulos,GLW}.  In particular 
\cite{GibbonsPapadopoulos} contains a complementary treatment of some
of the material in this section.

Let $G\subset\Spin_{10}$ and suppose we are given a configuration of
$G$-related planes.  As proven above, such a configuration preserves
at least a fraction $\nu_G$ of the supersymmetry.  This means that
there are at least $32\nu_G$ spinors $\varepsilon_i$ which obey
\begin{equation*}
(e_0\wedge\xi) \cdot \varepsilon_i = \varepsilon_i~,
\end{equation*}
for every oriented $5$-plane $\xi$ in the configuration.  As shown in
\cite{AFS-cali}, this means that every such $5$-plane $\xi$ is
calibrated by a (constant coefficient) $5$-form $\Omega_i$ in
$\EE^{10}$ which can be obtained from $\varepsilon_i$ by squaring.
Every such form $\Omega_i$ defines a face of the grassmannian of
oriented $5$-planes in $\EE^{10}$, known as the
$\Omega_i$-grassmannian.  By the Theorem, the subset of $G$-related
planes containing the given configuration is itself contained in (and
conjecturally agrees with) the intersection of the
$\Omega_i$-grassmannians.  As we will see in many examples below, this
subset often turns out to be itself isomorphic to the
$\Omega$-grass\-man\-nian for some ($p\leq5$)-form $\Omega$.
Therefore it defines a geometry in the sense of Harvey \& Lawson
\cite{HarveyLawson}.

As explained for example in \cite{Morgan-monthly}, a $p$-submanifold
(with possible self-intersections) of $\EE^{10}$ whose tangent spaces
lie in the same $\Omega$-grass\-man\-nian, is homologically
volume-minimising. In other words, the geometry associated with the
$\Omega$-grass\-man\-nian corresponds to the geometry of minimal
$p$-dimensional immersions in some euclidean space $\EE^{D\leq10}$.
Given $G\subset\Spin_{10}$, one determines $D$ and $p$ as follows:
$10-D$ is the dimension of the $G$-invariant subspace $V^G$ of
$\EE^{10}$, whereas $5-p$ is the dimension of subspace $V^G \cap
\xi$.  Equivalently, $p$ is equal to the codimension $d$ of the
configuration.  We have not classified all possible $G$, but we have
managed to construct a number of examples, which are summarised in
Table \ref{tab:geometries}.  This table refines Table
\ref{tab:summary} in that conjugate subgroups of $\Spin_{10}$, while
preserving the same fraction of the supersymmetry,  can give rise to
different geometries.

\begin{table}[h!]
\centering
\setlength{\extrarowheight}{5pt}
\begin{tabular}{|*{4}{>{$}c<{$}|}}
\hline
\multicolumn{1}{|c|}{Codim.}&
\multicolumn{1}{c|}{Group}&
\multicolumn{1}{c|}{Isotropy}&
\multicolumn{1}{c|}{Geometry}\\
d&
G&
K&
G/K\\
\hline\hline
5&\SU_5 & \SO_5 & \slg_5\\
&\SU_2\times\SU_3 & \SO_2\times\SO_3& (\CC_1~\text{or}~\slg_2)
\times\slg_3\\
\hline
&\Spin_7 &(\SU_2)^3/\ZZ_2 & \text{Cayley} \\
&\SU_4 & \SO_4 & \slg_4 \\
&\SU_4 &\mathrm{S}(\U_2\times\U_2) & \CC_2\\
&\Sp_2 &\U_2 & \clg_2 \\
4&\Sp_2 &\Sp_1\times\Sp_1 & \HH_1 \\
&\Sp_1\times\Sp_1 & \U_1\times\U_1 &
\CC_1\times\CC_1 \\
&\Sp_1\times\Sp_1 & \Sp_1 & (3,1)~\text{in
\cite{DadokHarveyMorgan}}\\
&\Sp_1 &\U_1 & (3,2)~\text{in
\cite{DadokHarveyMorgan}}\\
&\U_1 &\{1\} & (3,3)~\text{in
\cite{DadokHarveyMorgan}}\\[5pt]
\hline
3&G_2 &\SO_4 & \text{Associative} \\
&\SU_3 &\SO_3 & \slg_3\\[5pt]
\hline
2&\SU_3 & \mathrm{S}(\U_2\times\U_1) & \CC_1\\
&\SU_2 &\SO_2 &\CC_1~\text{or}~\slg_2\\[5pt]
\hline
\end{tabular}
\vspace{8pt}
\caption{Geometries associated with intersecting brane configurations.
Listed are some groups $G$ leaving spinors invariant, and the isotropy
subgroup $K\subset G$ which leaves the 5-plane $\xi$ invariant.  The
geometry of the resulting grassmannian $G/K$ is also listed.}
\label{tab:geometries}
\end{table}

Let us comment briefly on these results.  These examples have been
arrived at by choosing a convenient reference $5$-plane $\xi$ and
picking a number of linearly independent spinors in $\Delta(\xi)$.
The intersection $G$ of their isotropy subgroups inside $\Spin_{10}
\subset \Cl_{1,10}$ can be computed.  From this it is a simple matter
to determine the intersection $K$ of $G$ with the isotropy subgroup of
$\xi$.  Many of the calculations have been performed infinitesimally
(i.e., using their Lie algebras) using {\em
Mathematica\/}.\footnote{Details of the calculations can be obtained
by email from the authors.  They will be made public via our web pages
at a later date.}

We have included in the table only those groups $G$ for which we could
determine the geometry.  In particular, some groups in Table
\ref{tab:summary} associated with configurations with $d=5$ are
missing.  This reflects the present knowledge about the faces of the
grassmannian of oriented $5$-planes in $\EE^{10}$.  The determination
of the faces of the grassmannian $G(p,\EE^D)$ of oriented $p$-planes
in $\EE^D$ is not an easy problem whenever $p$ is different from
$1$, $2$, $D-2$, or $D-1$.  To this day, only the cases $(p,D) =
(3,6)$ \cite{DadokHarvey-R6,HarveyMorgan-6,Morgan-ext} and $(3,7)$
\cite{HarveyMorgan-7,Morgan-ext} have been fully solved, whereas there
are some partial results for $(p,8)$ \cite{DadokHarveyMorgan}.

Some of the geometries in the table are reasonably well-known: the
geometries of $p$-dimensional complex ($\CC_p$) or quaternionic
($\HH_p$) submanifolds are classical.  The special lagrangian
($\slg_p$), Cayley and associative geometries were discovered by Harvey
and Lawson in their foundational essay \cite{HarveyLawson}, and have
been discussed  recently in the context of intersecting branes in
\cite{GibbonsPapadopoulos,GLW,AFS-cali}.  Less known perhaps are the
complex lagrangian ($\clg_p$) geometry of $p$-dimensional complex
submanifolds in $\CC^{2p}$ which are lagrangian relative to a complex
symplectic form, and the geometries of types $(3,1)$, $(3,2)$ and
$(3,3)$.  These geometries are associated to faces of the grassmannian
$G(4,\EE^8)$ of oriented $4$-planes in $\EE^8$ which are calibrated by
self-dual $4$-forms.  They are discussed, together with explicit
representative calibrations, in \cite{DadokHarveyMorgan}.

Finally we should mention that there are more faces in the
grassmannian than the ones discussed here: we have only discussed
those faces which contain tangent spaces to supersymmetric brane
configurations.  The other faces correspond to cycles which
are not supersymmetric, yet are still minimal.  It may be interesting
to study these faces in more detail, particularly in the context of
Kaluza--Klein supergravity duality.

\section{Conclusions and open problems}

In this paper we have outlined a complete characterisation of
configurations of multiply intersecting branes at angles in terms of
subgroups of $\Spin_{10}$ preserving some spinors.  We believe that
this framework might be useful in the algebraic approach to intersecting
branes and in principle reduces the classification of such
configurations to a problem in group theory, which is roughly speaking
the decomposition of the grassmannian of oriented 5-planes in ten
dimensions in terms of orbits of subgroups $G$ of $\Spin_{10}$
contained in the isotropy of some spinor.  This approach suggests some
open problems.

The obvious open problem is settling the Conjecture, but there are
other problems as well.  One should do a systematic search of
subgroups of $\SU_5$ and $\Spin_7$ (i.e., of subgroups of $\Spin_{10}$
which are contained in the isotropy of a spinor) and determine the
fraction $\nu_G$ for them; maybe one finds fractions which are not
listed here.  We are not aware of any completeness result.  The groups
and fractions discussed in this paper are only 
complete for the case of two intersecting branes \cite{OhtaTownsend}.
Other groups, maybe even finite groups, may appear when one considers
more than two branes.  Each such group determines a `geometry' in the
sense of \cite{HarveyLawson}.  In other words, the orbit under this
group of the original $\M5$-brane defines a subset of the grassmannian
of oriented planes which, as explained at the end of the previous
section, can be associated with a certain geometry.  It would be
interesting to classify these geometries.  This is a refinement of the
(unsolved) problem of determining the faces of the grassmannian of
oriented 5-planes in $\EE^{10}$, since one need only consider those
faces which are intersections of the faces exposed by calibrations
which can be obtained by squaring spinors.

Another obvious problem, which will be addressed in Part~II is to lift
the restriction on the types of transformations one is allowed to do
on the branes.  We have followed \cite{OhtaTownsend} and allowed the
branes to be merely rotated relative to each other; but in fact, one
should allow for general eleven-dimensional Lorentz transformations.
A similar analysis is possible and one can classify all the
supersymmetric configurations involving only two branes as well as
prove some partial results for the case of an arbitrary number of
branes \cite{AFSS}.  No new fractions seem to emerge in this case
either.

There are other interesting aspects of intersecting $\M$-branes which
we have not addressed in this paper and for which this approach may be
fruitful. The duality between intersecting branes and Kaluza--Klein
supergravity \cite{GGPT} should be studied further, as are the
supergravity solutions corresponding to these more general
configurations.  We hope to report on these problems in the future.

\section*{Acknowledgements}

It is a pleasure to thank Robert Bryant for his helpful correspondence
and for sending us his unpublished notes \cite{Bryant-spinors}, and
Sonia Stanciu for useful discussions.  In addition, JMF would like to
thank Takashi Kimura and the Department of Mathematics of Boston
University for the hospitality during part of the time it took to
write this paper.  BSA is supported by a PPARC Postdoctoral
Fellowship, JMF by an EPSRC PDRA and BS by an EPSRC Advanced
Fellowship, and we would like to extend our thanks to the relevant
research councils for their support.

\end{document}